\documentclass{aa}

\usepackage{graphicx,txfonts,textcomp} 
\usepackage{natbib}
\usepackage[colorlinks=true,linkcolor=blue,citecolor=blue,urlcolor=blue]{hyperref}
\usepackage{enumitem}
\usepackage{booktabs}

\title{AGN_eJWST}
\author{Lenk et al.}
\date{May 2024}

\usepackage{listings} 

\usepackage{tabularx} 
\usepackage{orcidlink}

\definecolor{sqlkeyword}{rgb}{0.58, 0, 0.83} 
\definecolor{background}{rgb}{0.95, 0.95, 0.92} 

\lstdefinelanguage{SQL}{
    morekeywords={SELECT, FROM, WHERE, AND, OR, IS, NOT, NULL}, 
    keywordstyle=\color{sqlkeyword}\bfseries, 
    identifierstyle=, 
    commentstyle=, 
    stringstyle=, 
    basicstyle=\ttfamily, 
    backgroundcolor=\color{background}, 
    showstringspaces=false, 
}

\begin{document}

   \title{The eJWST active galactic nucleus observation catalogue}


   \author{V. Lenk\orcidlink{0009-0000-6547-7959} \inst{1,2},
          A. Labiano\orcidlink{0000-0002-0690-8824}\inst{3}
          \and
          C. Circosta\orcidlink{0000-0001-8522-9434}\inst{2,4}
          \and
          A. Alonso-Herrero\orcidlink{0000-0001-6794-2519} \inst{5}
          \and
         D. Wylezalek\orcidlink{0000-0003-2212-6045}\inst{1}
          }

   \institute{Zentrum für Astronomie der Universität Heidelberg, Astronomisches Rechen-Institut Mönchhofstr, 12-14 69120 Heidelberg, Germany. 
            \email{virginialenk@gmx.de}
             \and
             European Space Agency (ESA), European Space Astronomy Centre (ESAC), Camino Bajo del Castillo s/n, 28692 Villanueva de la Ca\~nada, Madrid, Spain.
             \and
             Telespazio UK S.L. for ESA, ESAC, Camino Bajo del Castillo s/n, 28692 Villanueva de la Ca\~nada, Madrid, Spain.
             \and
             Institut de Radioastronomie Millimétrique (IRAM), 300 rue de la Piscine, 38400 Saint-Martin-d’Hères, France.
             \and
             Centro de Astrobiolog\'{\i}a (CAB), CSIC-INTA, Camino Bajo del Castillo s/n, 28692 Villanueva de la Ca\~nada, Spain.
             }

   \date{Received ..., 2025; accepted ..., 2025}


  \abstract
   {The European Archive of the \textit{James Webb} Space Telescope (eJWST) provides access to all data collected by the \textit{James Webb} Space Telescope (JWST). JWST's capabilities span from studying early Universe galaxy formation to probing exoplanet atmospheres. Specifically, for active galactic nuclei (AGNs), JWST offers unparalleled opportunities, enabling investigation into AGN phenomena with unprecedented detail through high-resolution imaging, spectroscopy, and photometric data.}
   {This study aims to compile and release a catalogue of all AGN observations conducted with JWST. Using eJWST, we systematically filtered and organized these observations to facilitate access and retrieval of all of JWST's data products related to AGNs. Our goal is to provide the community with a valuable resource for their research.
   }
   {We compiled the AGN observations in eJWST using specific keywords set by the principal investigators in their proposals, manually reviewing the approved programmes of JWST, as well as cross-matching all available observations with available AGN catalogues, such as the Million Quasar catalogue, the SDSS MaNGA AGN catalogue, the CDFS catalogue, and others.
   }
   {The resulting catalogue contains a total of 3,242 individual AGNs included in JWST observations. This is one of the first extensive collections of AGN observations from the JWST. It includes detailed information about the targets (name, coordinates, and redshift) and specifics of the JWST observations (instrument, aperture, filter, etc.), and provides links for data downloads.}
  
   {}

    \titlerunning{The eJWST AGN observation catalogue}

    \authorrunning{V. Lenk et al.}

   \maketitle
%

\section{Introduction}
The \textit{James Webb} Space Telescope (JWST), launched in December 2021, is a space infrared observatory built by the National Aeronautics and Space Administration (NASA), the European Space Agency (ESA), and the Canadian Space Agency (CSA). It represents a monumental advancement in astronomical observation, providing extraordinary sensitivity and wavelength coverage from 0.6 to 28 microns  \citep{Gardner23, Rigby23}. JWST is equipped with four scientific instruments: the Near Infrared Camera \citep[NIRCam;][]{Rieke2005, Rieke2023}, the Near Infrared Spectrograph \citep[NIRSpec;][]{Ferruit2012, Boker2023}, the Mid-Infrared Instrument \citep[MIRI;][]{Rieke2015, Wright2015, Wright2023}, and the Fine Guidance Sensor/Near InfraRed Imager and Slitless Spectrograph \citep[FGS/NIRISS;][]{Doyon2012, Doyon2023}.
The telescope transmits $\gtrsim$ 28.6 Gb of data to Earth twice per day\footnote{\url{https://jwst-docs.stsci.edu/jwst-general-support} and \citet{Gardner23}}. 
These data are then sent to the Deep Space Network (DSN), a global network of large radio antennas located in Goldstone, California, Madrid, Spain, and Canberra, Australia.
After the accumulation of data on the ground, they are transmitted to the Space Telescope Science Institute (STScI) in Baltimore \citep{Swade2014, Greene2015}. There, the Data Management System (DMS) processes them to three different calibration levels: Calibration Level 1 is the initial stage of the JWST data pipeline. At this level, basic calibration corrections for instrumental effects, such as dark current or the flagging of known bad pixels, are applied. The output consists of count-rate images for each exposure.  Calibration Level 2 is the second stage of the pipeline, where further processing of Level 1 data occurs. This includes pixel flat-fielding, derivation, and attachment of world-coordinate information. The output at this stage includes fully calibrated data from individual exposures. The final stage of the pipeline is Calibration Level 3, where individual exposures of Calibration Level 2 are joined to science-ready mosaics, integral field unit (IFU) data cubes, or one-dimensional spectra \citep{Bushouse2020}. Planned observations that have been approved but not yet executed are not part of the calibration pipeline, but they are listed in the archives under Calibration Level -1. 

The DMS uses software pipelines designed to handle each of the various modes of operation of the science instruments on board the telescope. These pipelines are frequently updated as the instrument's behaviour, calibration, and knowledge evolve. At the time of writing, the latest available JWST pipeline is 1.17.0 \citep{Bushouse2024}.

\begin{table*}[h!]
\caption{AGN descriptions available in APT for Cycle 4 proposals.}
\label{AGN descriptions}
\centering
\begin{tabularx}{\textwidth}{X X X}
Quasars & Active galactic nuclei & Radio jets \\
Double quasars & Active galaxies & Radio cores\\
Radio loud quasars & Seyfert galaxies & Radio lobes \\
Radio quiet quasars & Blazars & Radio galaxies \\
X-ray quasars & LINER galaxies &  Radio plumes \\
Broad-absorption line quasar & Galactic accretion disk & Radio hot spots \\
Quasar-galaxy pairs & Ultraluminous infrared galaxies & Galaxy jets \\
Relativistic jets &  BL Lacertae objects &  \\
\end{tabularx}
\end{table*}

After processing, the data are archived in long-term storage systems (archives) and made available to researchers. The three agencies involved in the development of the mission host JWST data archives with different functionalities: the European archive of the \textit{James Webb} Space Telescope \citep[eJWST\footnote{\url{https://jwst.esac.esa.int/archive}}; e.g.][]{Labiano2023,Arevalo2024} at the European Space Astronomy Centre (ESAC), the Mikulski Astronomical Space Telescope (MAST) at the STScI, and the JWST archive at the Canadian Astronomical Data Centre (CADC). The three archives use the same common archive observation model \citep[CAOM;][]{Dowler2012} to read and access JWST metadata, so identical versions of every file are hosted on the three sites.

The eJWST is designed to be a service to the European scientific community but is open to all scientists around the world, enabling direct and efficient access to all the JWST science data products. The eJWST forms part of ESA's science archives. It is therefore searchable in combination with data from any other ESA science mission. The eJWST holdings consist of all public data from the JWST mission, as well as access to all proprietary data. The eJWST provides multiple methods for searching for and filtering products and calibration data, such as downloading directly from the user interface or using Astronomical Data Query Language \citep[ADQL;][]{Mantelet2023}, Astroquery \citep{Ginsburg2019}, CURL, Python, or WGET queries.

One of the primary goals for the JWST is to study galaxy evolution over time. In this context, active galactic nuclei (AGNs) play a crucial role by influencing the evolution of different properties of their host galaxy \citep{Graham2011, Wagner2012, Silk2013, Kormendy2013, Harrison2017, Cresci2018, Circosta2021}. 
The JWST observations have opened a new window for AGN studies, offering significant advances in resolution and sensitivity at infrared wavelengths, and enable the detection of distant and faint AGNs or AGNs that were previously undetected \citep[e.g. too faint or obscured by dust;][]{Harikane2023,Lyu2024}. Over the years, the list of AGNs has increased steadily, as shown by the releases of numerous catalogues such as the the Two-degree Field Galaxy Redshift Survey (2dF) catalogue \citep{Croom2001}, the Sloan Digital Sky Survey \citep[SDSS;][]{Abazajian2003}, and the 7Ms catalogue of the \textit{Chandra} Deep Field-South Survey \citep[CDFS;][]{Luo2017}, which resulted in large data collections of AGNs.
Creating a comprehensive catalogue of AGNs observed by JWST will be helpful for scientists aiming to explore the impact of nuclear activity on galaxy evolution.

Here we present a catalogue of all the JWST observations of AGNs to date. 
To create the list of AGNs, we filtered observations directly described as AGNs by the proposers and cross-matched
AGN catalogues from the literature with all data in eJWST to provide the community with a comprehensive list of targets and observation modes used to observe them. In Sect. \ref{data} we give a brief introduction on how to access data stored in the eJWST and how to extract datasets. In Sect. \ref{agn selection} we describe the search and filtering processes used to create the catalogue. 
For the redshift cross-match in this section, we assume a flat $\Lambda$ cold dark matter cosmology with $H_0 = 70\text{ km s}^{-1}\text{Mpc}^{-1}, \Omega_m =0.3$, and  $\Omega_\Lambda =0.7$.
In Sect. \ref{quality} we describe the different caveats identified throughout the process. 
In Sect. \ref{ambigious AGNs} we discuss objects whose statuses are currently under debate and ongoing surveys that have not yet been included. Then we give an overview of the catalogue and its content in Sect. \ref{catalogue description}, and recap our work in Sect. \ref{conclusion}.

\section{Data access in eJWST}
\label{data}
There are six different types of JWST observational programmes: the General Observer (GO) programme, the Guaranteed Time Observations programme, the Director's Discretionary Time (DDT), the Director’s Discretionary Early Release Science (DD-ERS), the Calibration programmes (which include commissioning data), and the Early Release Observations. Each observation from these programmes is accompanied by data products such as high-resolution images, spectra, or time-series observations, all made publicly available at the JWST archives.
The eJWST can be accessed either through the web interface or through a TAP+ REST service \citep{TAP2019}. Through the TAP+ service users can query metadata and datasets using the ADQL \citep{Mantelet2023}.

In this work we made use of this service through the \verb|astroquery.esa.jwst| package in Python, which allowed us to perform ADQL queries (example in Listing \ref{lst:crossmatch_adql}) to access the metadata directly from eJWST.
The eJWST provides several different CAOM tables, storing raw data as well as various calibration levels of observation data. Predominantly, the \verb|jwst.observation| table (the main CAOM observation table) 
and the \verb|jwst.archive| table (housing data visible on the eJWST archive) were used for this catalogue. From the \verb|jwst.observation| table, we extracted mainly keywords storing instrument (\verb|instrument_keywords|) and target (\verb|target_keywords|) information, while the \verb|jwst.archive| table was used to extract information on the observation itself, such as the process level (\verb|calibrationlevel|), availability to the public (\verb|public|), and metadata release date (\verb|metarelease|). This was necessary because some keywords were exclusive to one table. 
Upon query execution, data were obtained in the form of an \verb|astropy.table| object, subsequently converted into a \verb|PandasDataframe| using the Python package \verb|Pandas| \citep{Pandas24}. 

\section{AGN selection in eJWST}
\label{agn selection}
The goal of this work is to compile a catalogue of all observations with JWST where an AGN is present in the field of view (FoV). 
Therefore, it is necessary to include not only data from proposals specifically targeting AGNs (throughout the paper we refer to them as `targeted AGNs'), but also any data where an AGN was within the FoV regardless of the primary objective of the observation (`non-targeted AGNs'). This includes not only science observations, but also calibration and background data.

All archive searches incorporated both public and private science-ready products (calibration level 3) as well as proposed observations (calibration level -1).  
To ensure that all AGN observations, whether targeted or not, were included in the catalogue, three different methods were applied:
\begin{itemize}
    \item Search by keywords in the proposals.
    \item Search within approved programmes.
    \item Cross-match AGN catalogues with the FoVs of individual JWST observations.
\end{itemize}
The first two methods (Sects. \ref{keywords} and \ref{approved program}) compile a list of all targeted AGNs, while the third method (Sect. \ref{cross-matching}) includes both targeted and non-targeted AGNs. The three methods were then cross-checked to find matches within different catalogues and to eliminate duplicated targets (see Sect. \ref{quality}). The following subsections describe each method in detail.

\subsection{By keywords from proposals}
\label{keywords}
\begin{figure}[ht]
    \centering
    \includegraphics[width=\columnwidth, trim={0 10pt 150pt 0}, clip]{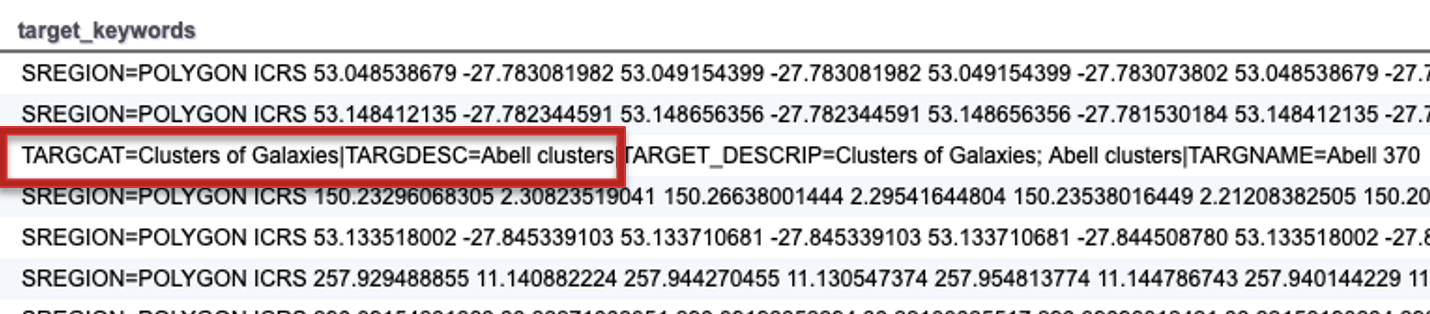}
    \caption{Snippet of target\_keywords with the category and description label highlighted.}
    \label{fig: Keywords set by proposers.}
\end{figure}

The Astronomer's Proposal Tool (APT)\footnote{\url{https://www.stsci.edu/scientific-community/software/astronomers-proposal-tool-apt}} has the option of adding a category and description label to the targets in the proposal. These labels are stored in the CAOM tables.  
They are accessible through the keyword \verb|target_keywords|, which stores target-specific information separated by a vertical bar `|' (see Fig. \ref{fig: Keywords set by proposers.}). Here, the target category and description are found in two versions: the labels \verb|TARGCAT| and \verb|TARGDESC|  are used for already observed objects, and TARGETCATEGORY| and  \verb|TARGETKEYWORDS| for proposed objects (calibration level -1). The \verb|TARGCAT| and \verb|TARGETCATEGORY| labels describe the category of the target (e.g. galaxy, star, etc.), each target being assigned a single category. \verb|TARGDESC| and \verb|TARGETKEYWORDS| provide the target description (e.g. AGNs, protostars, etc.), with the potential for multiple descriptions per target. 

We selected all AGN descriptors (see Table \ref{AGN descriptions}) from the category `Galaxy' in the proposal parameters documentation\footnote{\url{https://jwst-docs.stsci.edu/jppom/targets/fixed-targets/target-descriptions\#gsc.tab=0}} to search for all targeted AGNs in the eJWST.  
It is important to acknowledge that some targets lack a description entirely (see Sect. \ref{approved program}) especially in early data (commissioning and Cycle 1 mainly). These targets lack the descriptor not only in the CAOM tables but also in the flexible image transport system files, and were thus missed by this search method. To complete the sample with these targets, we used the method described in Sect. \ref{approved program}.

The target description and categories used for filtering 
may need to be updated in future catalogue releases if new target descriptions or categories are added to the APT in the next Calls for Proposals. We found 901 individual targeted AGNs using this method.

\subsection{By approved programmes}
\label{approved program}
The target descriptions retrieved from the aforementioned keywords are dependent on the proposers' input. As a result, we could not assume that targets were consistently labelled as AGNs, especially since sometimes targets lacked the description field. Also, some AGN sources can fall under different descriptors (e.g. starburst or high-redshift galaxy) and are not labelled as AGNs in the APT field. 
To address possibly missed AGNs, we implemented an additional search. We searched the approved STScI JWST listed programmes\footnote{\url{https://www.stsci.edu/jwst/science-execution/approved-programs}} for additional AGNs, described either in the abstract or elsewhere in the proposal. Specifically, we checked the GO and the Guaranteed Time Observation programmes during all available cycles, as well as the ERS programmes, the DDT, and the calibration programmes. From the GO we checked all the proposals listed under `Galaxies' and `Supermassive Black Holes and Active Galaxies'. 
Naturally, some of these proposals include targets already found in our previous search. We cross-checked the two searches and removed duplicated results. This search method increased our sample to 968 targeted AGNs.

\begin{figure*}[htbp]
    \centering
    \begin{lstlisting}[language=SQL, caption={ADQL query for cross-match.}, label={lst:crossmatch_adql}]
SELECT * 
FROM jwst.archive 
WHERE ((jwst.archive.calibrationlevel = 3) OR (jwst.archive.calibrationlevel = -1)) 
AND (position_bounds_spoly IS NOT NULL 
AND INTERSECTS(CIRCLE('ICRS',ra,dec,0),position_bounds_spoly)=1)
\end{lstlisting}
\end{figure*}

\subsection{Cross-matching with AGN catalogues}
\label{cross-matching}
The high sensitivity of JWST, and the large FoV of some of its instruments, make it possible to detect serendipitous AGNs in the FoV of any observation, regardless of the primary objective of the data (science, calibration, background, etc) or scientific goal of the observation. To build the most complete catalogue of sources we need to include those serendipitous, non-targeted AGNs in the catalogue. The coordinates describing the FoV for any JWST observation are stored in the \verb|position_bounds_spoly| column of the \verb|jwst.archive| table. This column contains four coordinate pairs (RA, Dec) that outline a polygon in the sky. 

To account for these non-targeted sources, we cross-matched all targets from various AGN catalogues in the literature (see Sect. \ref{Sample}) with the polygons representing the FoVs of all JWST observations.
The eJWST interface offers a versatile and user-friendly platform for downloading tables in multiple formats, including VOTable, comma-separated values, and the flexible image transport system. By utilizing these download options, we retrieved the JWST observations through an ADQL query directly on the interface without incurring significant time costs. 

After retrieving all JWST observations, we cross-matched them with the target coordinates from the AGN catalogues using TOPCAT's \citep{Taylor2005} \verb|inSkyPolygon| function, to check if the coordinates fell within any polygon (see Fig. \ref{fig:masked_area2}).
TOPCAT’s \verb|inSkyPolygon| tool provides a practical approach for spatial querying within a specified polygonal region of the sky. The immediate and interactive nature of \verb|inSkyPolygon| is particularly useful when dealing with large datasets that need filtering based on spatial location. 

A similar approach is offered by the eJWST interface directly, using ADQL queries (see Listing \ref{lst:crossmatch_adql}). Here, an intersect function is used to verify if a circle with a certain radius around the target coordinates lies within the polygon. For the matching procedure, we did not want to use a search radius around the AGN coordinates to avoid imposing arbitrary constraints, such as defining the matching area as a fraction of the FoV, or deciding which coordinates to prioritize. Also, this could cause false positives in regions close to the edges of the FoV. Thus, we carried out a cross-match to pixel coordinates instead of the radius search.  To do this in the eJWST, we set the radius to 0. This yields the same result as the \verb|inSkyPolygon| function but, due to its being dependent on the database server load, filtering large datasets (such as those with $\geq$600 entries) can be computationally expensive and time-consuming. Therefore, the on-the-fly filtering of \verb|inSkyPolygon| aligned better with our requirements for fast and efficient data analysis in this context.

\subsection{Catalogues}
\label{Sample}
To identify AGNs outside of the JWST proposals, we utilized the following key catalogues and literature sources to ensure comprehensive coverage and accuracy. The main properties of all catalogues used can also be found in Table \ref{tab:catalog_summary}.
\begin{itemize}[label=--]
    \item The {Million Quasar} (Milliquas) catalogue by \cite{Flesch2023} includes all published quasars up to June 30, 2023, featuring 907,144 type-I quasi-stellar objects and AGNs, 66,026 high-confidence radio/X-ray associated candidates, and 48,630 type-II and BL Lac objects, totaling 1,021,800 entries. Continuously updated since 2009, the catalogue maintains records from various sources, including data from the Dark Energy Spectroscopic Instrument (DESI), the Sloan Digital Sky Survey Data Release 18 (SDSS-DR18), the Second XMM-Newton Serendipitous Source Catalogue (2XMM-Newton), the Cosmic Evolution Survey (COSMOS), the Wide-field Infrared Survey Explorer (WISE), the 2dF , the Six-degree Field Galaxy Survey (6dF), and the Two Micron All Sky Survey (2MASS).
    By cross-matching the Milliquas catalogue with all JWST observations, we find 2,495 AGNs from the Milliquas catalogue in one or more JWST observations.
    \item The {SDSS MaNGA AGN} (MaNGA) catalogue by \cite{Comerford2024} identifies AGNs in the final DR17 MaNGA sample of approximately 10,000 nearby galaxies using mid-infrared, X-ray, and radio data, alongside broad emission lines in SDSS spectra. hey used data from the Wide-field Infrared Survey Explorer (WISE), the Burst Alert Telescope (BAT), the NRAO VLA Sky Survey (NVSS), and the Faint Images of the Radio Sky at Twenty Centimeters (FIRST), applying selection criteria in these wavelengths, and identified a total of 397 AGNs. The cross-matching resulted in multiple JWST observations for 7 of these AGNs.
    \item The {eROSITA Final Equatorial Depth Survey} (eFEDS) by \citet{Lui2022} is one of the largest contiguous-field X-ray surveys observing a total area of 142 $\text{deg}^2$. 
    It provides a comprehensive catalogue of X-ray sources with extensive multi-band photometric and spectroscopic coverage. The eFEDS AGN catalogue includes 22,079 AGNs from the 27,910 sources found by the main X-ray catalogue, predominantly featuring X-ray unobscured AGNs. The cross-matching found 8 of them in JWST observations.
    \item The {Stripe 82X survey} (Stripe) multi-wavelength catalogue by \citet{Ananna2019} covers 31.3 $\text{deg}^2$ using \textit{Chandra} and \textit{XMM-Newton}.  
    The updated catalogue includes 5,961 X-ray sources with multi-wavelength counterparts providing spectroscopic and photometric redshifts. We selected for our sample all objects listed as Type 1 and Type 2 AGNs, as well as quasi-stellar objects (QSOs), which resulted in a total of 2,459 AGNs. From these, 27 were found in JWST observations.
    \item The {CDFS: 7 Ms Source catalogues} \citep[][]{Luo2017} is an X-ray source catalogue covering 484.2 arcmin$^2$. The main catalogue includes 1,008 X-ray sources, providing object classification of 711 AGNs with mostly spectroscopic redshifts. Here, 419 of those AGNs were observed by JWST.
    \item The second data release of the {\textit{Swift}/BAT AGN Spectroscopic Survey} (BASS) presents a catalogue of AGNs by \citet{Koss2022} along with their optical spectroscopy. This release includes 858 hard-X-ray-selected AGNs from the \textit{Swift}/BAT 70-month sample, updating the previously released catalogue of AGNs by \citet{Baumgartner2013}. Using instruments like Very Large Telescope (VLT)/X-shooter and Palomar/Doublespec they offer high-resolution spectra and broad wavelength coverage (3200–10,000 Å). This release is notable for its completeness in spectroscopic coverage (100\% for AGNs), with nearly all AGNs having spectroscopic redshift measurements (99.8\%). We found 61 of them in JWST observations.
    \item The {catalogue of quasars and active nuclei: 13th edition} (Veroncat), published by \citet{VCV2010}, is the follow-up of the previously published catalogues \citep{VCV2006, VCV2003}. This edition primarily uses the fifth, sixth, and seventh data releases from the SDSS \citet{McCarthy2007, McCarthy2008, Abazajian2009}, presenting a compilation of 133,336 quasars, 1,374 BL Lac objects, and 34,231 active galaxies, including 16,517 Seyfert 1s. They include measured redshifts known before July 1, 2009, and positions. We made use of their AGN table (34,231 objects), which lists `active galaxies' such as Seyfert 1s (S1), Seyfert 2s (S2), and and Low-Ionization Nuclear Emission-line Regions (LINERs) as defined by \citet{Heckman1980}. In total 399 of these AGNs were found in JWST observations.
    \item The {ROSAT-FIRST AGN} catalogue (Rosat) by \citet{Brinkmann2000}, combines data from the ROSAT All-Sky Survey (RASS) and the Very Large Array (VLA) FIRST 20cm survey, focusing on 843 X-ray sources with unique radio counterparts. The catalogue uses RASS for X-ray data and FIRST for high-sensitivity radio data (1 mJy at 1.4 GHz) with accurate positioning (about 1 arcsecond). In total 294 confirmed AGNs
    are identified. Here, 5 AGNs were found in JWST observations.
    \item The {\textit{Gaia}-Celestial Reference Frame 3} (\textit{Gaia}-CRF3) \citep{Gaia2022} filters QSO/quasar targets from the \textit{Gaia} Data Release 3 (\textit{Gaia} DR3) with over 90\% purity. The \textit{Gaia}-CRF3 content lists QSO-like sources retrieved from 17 external catalogues, for example the All Wide-field Infrared Survey Explorer (AllWISE) (\citeauthor{Secrest2015} \citeyear{Secrest2015}), Milliquas v6.5 (\citeauthor{Flesch2015} \citeyear{Flesch2015}), R90  (\citeauthor{Assef2018} \citeyear{Assef2018}), the Fifth Large Quasar Astrometric Catalog (LQAC-5) (\citeauthor{Souchay2019} \citeyear{Souchay2019}), and more. Accessing the \verb|gaiadr3.qso_candidates| table via the \textit{Gaia} archive\footnote{\url{https://gea.esac.esa.int/archive/}}, we obtained a dataset that contains extragalactic sources but has low purity due to a completeness-driven selection. To then select the \textit{Gaia}-CRF3 targets, we applied the criterion \verb|gaia_crf_source = 'True'|, which retrieves all 1,614,173 sources from the CRF3 table. To align with the JWST coordinates epoch, we directly transformed the given coordinates from the J2016 to the J2000 epoch using the \textit{Gaia} archive's \verb|ESDC_EPOCH_PROP| function. In total 400 AGNs were found in one or several JWST observations.
    \item The {local \textit{Swift}/BAT AGN} (Swiftbat) study provides a list of 277 local hosts of selected AGNs from the 58-month \textit{Swift}/BAT AGN sample \citep{Lutz2018L}. Covering $14-195$ keV X-rays they ensure detection of all but the most Compton-thick AGNs, including only z<0.06 objects. 
    The study considers only confidently identified AGNs, excluding low-luminosity AGNs below the BAT detection threshold. The cross-matching revealed that 44 AGNs from this sample were found in one or several JWST observations.
\end{itemize}

\begin{figure}[t]
    \centering
\includegraphics[width=1.0\columnwidth]{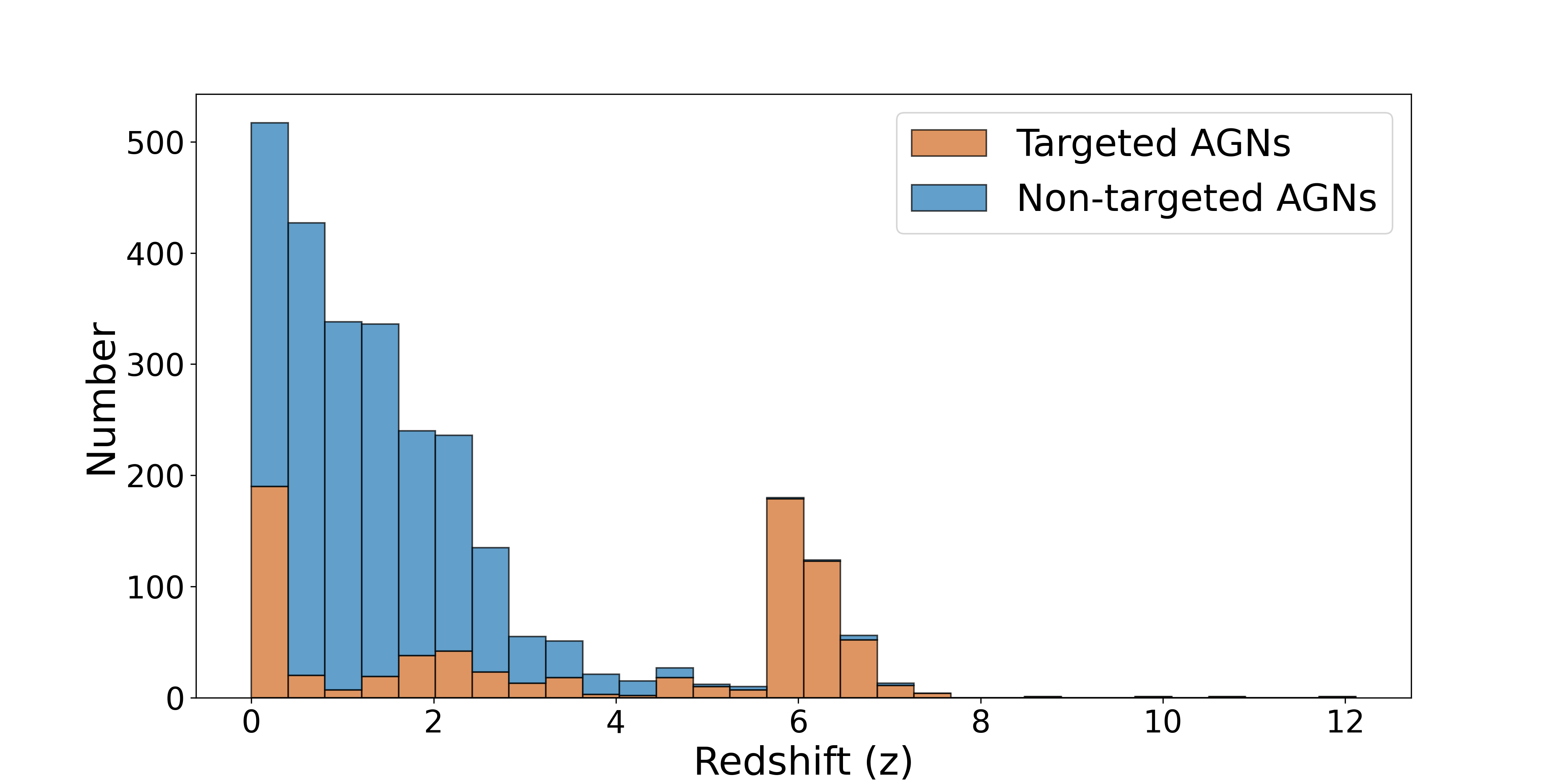}
    \caption{Histogram of redshifts in the eJWST AGN catalogue. 
    See text for details.}
    \label{fig:redshift}
\end{figure} 

\subsection{Redshift retrieval}
\label{redshift}
When redshifts were available in the external catalogues, we extracted and stored them in separate columns for each catalogue, for example `z (catalogue name)'. If a catalogue provided both spectroscopic and photometric redshifts, the spectroscopic redshifts were prioritized.

When available, information on the redshift type (spectroscopic or photometric) and the corresponding reference (in the form of a bibcode) was extracted from each catalogue or, when necessary, from the associated catalogue paper. This information was then stored in dedicated columns, for example `zType (catalogue name)'. If the catalogue or its description explicitly stated the method used, we assigned the redshift type accordingly (e.g. identification by spectral lines → spectroscopic). If no clear information was available, the type was set to `?'. For catalogues providing their own redshifts (e.g. MaNGA, eFEDs, and Stripe), the bibcode of the corresponding publication was used as reference, whereas for large compilations of literature values (e.g. Milliquas, Véron-Cetty, and Véron), the reference columns given by the catalogue were used to identify and assign the appropriate bibcodes.

Since many of the targeted AGNs lacked redshift information in either the proposal or catalogue data, we opted to match them with the 
NASA/IPAC Extragalactic Database (NED). Using the \verb|query_region()| function from the \verb|astroquery.ned| package, we performed a coordinate-based matching procedure with a 1-arcsecond search radius, using the target coordinates provided by eJWST for all objects in our catalogue. 
This coordinate-based query was used to return the redshift, redshift type and redshift reference from NED. We then compared NED redshifts with catalogue values and removed NED matches when the difference exceeded 0.01. For the remaining reliable matches, the NED results were stored in dedicated columns: z (NED),  zType (NED), and zRef (NED).  
To integrate these results into the working dataset, the NED values were merged into the main redshift (z), redshift type (zType), and redshift reference (zRef) columns wherever available. Redshift types were assigned according to the NED classification scheme, as outlined in their database guide\footnote{\url{https://ned.ipac.caltech.edu/Documents/Guides/Database}}. In cases where no NED entry was returned, we kept the catalogue-provided values according to the following priority order: Milliquas, MaNGA, eFEDS, Stripe, CDFS, BASS, Veroncat, and Rosat. This priority order was established based on the completeness, date of measurement, and accuracy of the information from each source.  Missing redshifts were marked as Not a Number (NaN).  To maintain transparency, we introduced a `zNED' flag. Entries with NED-derived redshifts are marked with `y', while those with catalogue-based values are marked with `n'.  
This framework ensures that catalogue redshifts are preserved while NED information is systematically incorporated and clearly identified.

In total we found and list redshifts for 86.78\% 
(2,814)
of all individual AGNs (3,242) in our catalogue. A distribution of the redshift for all AGNs (targeted and non-targeted) can be found in Fig. \ref{fig:redshift}.

\subsection{Object type retrieval}
\label{objtype}
For the targeted AGNs, the object classifications were retrieved directly from the descriptions provided by the proposers. For the non-targeted AGNs, we extracted the object types from the external catalogues whenever available. These catalogues differ in how they report AGN subclasses. For example, Veroncat distinguishes between type-1 and type-2 AGNs, and intermediate subclasses such as S1.0–S1.9 for intermediate Seyferts, S2 and S2? for confirmed and probable type-2 AGNs, S3 for LINERs, and BL for confirmed BL Lacs. While BASS specifies Seyfert subclasses (e.g. Sy1, Sy1.9, and Sy2) and additionally distinguishes beamed AGN types such as blazars (BZQ, BZG, and BZB).
To avoid ambiguity, we did not directly adopt catalogue abbreviations, since identical codes are sometimes used with different meanings across catalogues (e.g. `S' denoting either `Seyfert' or `Star'). Instead, we constructed a harmonized classification scheme based on the catalogue descriptions, writing out the types as clearly as possible while retaining minimal abbreviations. For example, the Milliquas classification `A = AGN, type-I Seyferts/host-dominated' was recast as `AGN type-1', and `K = NLQSO, type-II narrow-line core-dominated' was listed as `QSO-NL'. Where available, we also incorporated information on whether the source is narrow-line (NL) or broad-line (BL). A `?' was added to object types to indicate uncertain classifications (e.g. `AGN-BL?'), `??' for highly questionable assignments, and a single `?' (with no object type specification) was used for completely unknown types. Additional refinements were applied to preserve catalogue-specific information, such as `-b' in Veroncat for broad Balmer lines, `-h' for polarized Balmer lines, and `-i' for broad Paschen lines observed in the infrared.
A full description of the original classifications and the mapping is provided in Tables~\ref{tab:milliquas_map}--\ref{tab:stripe_map}.

\subsection{Identifying duplicates}
\label{macthingprocess}
\begin{table*}[ht!]
\centering
\caption{Example of matches found via redshift-dependent cross-match for the
same observation in two catalogues.}
\label{matches}
\resizebox{\textwidth}{!}{%
\begin{tabular}{llrrrrl}
\toprule
Milliquas Name & Veroncat Name & RA (Milliquas) & Dec (Milliquas) & RA (Veroncat) & Dec (Veroncat) & observationid \\
\midrule
Q J12367+6215 & Q J12367+6215 & 189.175967 & 62.262648 & 189.175833 & 62.262500 & jw01181-c1009\_s03562\_nirspec\_f290lp-g395m \\
BX 1335 & BX 1335 & 189.186667 & 62.258611 & 189.186250 & 62.258611 & jw01181-c1009\_s03562\_nirspec\_f290lp-g395m \\
Q 1234+6231 & NaN & 189.095554 & 62.257402 & NaN & NaN & jw01181-c1009\_s03562\_nirspec\_f290lp-g395m \\
A03.161 & NaN & 189.122696 & 62.253620 & NaN & NaN & jw01181-c1009\_s03562\_nirspec\_f290lp-g395m \\
HETDEX 2970 & NaN & 189.194290 & 62.246147 & NaN & NaN & jw01181-c1009\_s03562\_nirspec\_f290lp-g395m \\
NaN & B2.161 & NaN & NaN & 189.122500 & 62.253611 & jw01181-c1009\_s03562\_nirspec\_f290lp-g395m \\
\bottomrule
\end{tabular}%
}
\end{table*}
After the cross-matching process, results from each catalogue were combined into a single data frame using an outer merge (which returns all the rows from both data frames, with NaN for any missing matches) on the `observationid'. This groups all objects identified within the same FoV of observation together, regardless of whether they were the same objects. These groups were then analysed to identify cases where the same object was listed in multiple catalogues, ensuring that duplicates were not included.
Our approach involved two steps: first, a redshift-based coordinate cross-match between all targets found within the same observation. Using the redshift data provided by the catalogues, we calculated an appropriate matching radius by determining the angular separation in arcseconds that corresponds to a proper kiloparsec at the given redshift z for each object:
\begin{equation}
    \theta_{\text{arcsec}} = \frac{x_{\text{kpc}}}{D_A}, \text{ where } 
D_A = \frac{S_k(r)}{1+z}
.\end{equation}
Here $\theta_{\text{arcsec}}$ is the angular separation in arcsec, $x_{\text{kpc}}$ the physical size of the object in kpc, $D_A$ the angular diameter distance, z the redshift, and $S_k(r)$ the Friedmann–Lemaître–Robertson–Walker metric. We made use of the \verb|arcsec_per_kpc_proper(z)| function from the \verb|astropy.cosmology.funcs| module to determine $1/D_A$. 
The physical size was set to 1 kpc to ensure that only objects with very close angular positions are considered as the same. This leverages the high positional accuracy typically reported in the catalogues as well as taking into account the considerable fraction (14.35\%) of objects of redshifts between $5$ and $8$ (see Fig. \ref{fig:redshift}). Since galaxy sizes generally decrease with increasing redshift \citep{vanderWel2014, Whitney2019, Costantin2023, Ormerod2024}, a threshold of 1 kpc is reasonable. For example, \citet{Whitney2019} found that galaxies at $z=7$ have a median size of 1.64 kpc.  
The matches identified by this procedure were manually inspected, with particular attention to the redshifts (when available) to ensure the objects were not mistakenly matched due to projection effects. 
Secondly, we compared the given names of the AGNs, along with their redshifts, for the objects not previously found as matches. If the name (excluding spaces and hyphens `-') and the provided redshift were identical, the targets were considered the same. Each identified match was then subject to additional manual inspection, including verification against databases such as NED.
All objects without redshift information were also manually inspected, and, when possible, matches were verified through the use of NED. Only targets identified with high certainty as matches were noted as such; otherwise, they were retained as individual targets. 

After this matching procedure, we now count a total of 3,242 unique AGNs in our catalogue. With a total of 968 targeted and 2,274 non-targeted AGNs. Individual AGNs were counted by coordinates with a matching radius of 1 arcsecond. This counting method is kept throughout the work. For matches with the largest separation (>0.5 arcsec) individual checks were done and possible miscounted objects were corrected.

\subsection{Background observations}
We provide information on whether the main goal of an observation was science, calibration, or background. Data intended for science and calibration were filtered using the `Intend' keyword from the archive. To identify background data, we filtered by instrument keyword \verb|bkgdtarg = t|(rue). 

We also searched for the \verb|TARGNAME| keyword in the \verb|target_keywords| for specific descriptors identifying background observations, such as `background', `BCKGND', `Background', and more. Observations with these descriptors were then labelled as background observations.
Some targets were not explicitly labelled as background by such keywords but were identified as such by their target names. To address this, we conducted a further check for any names containing terms suggestive of background observations (e.g. NGC 6552-Background). All observations identified as background through this process were marked accordingly with the entry `Background' in the intent column. In total 1,150 observations were marked as background. 

\section{Caveats}
\label{quality}
\subsection{Accuracy of the (cross-)matching}
\begin{figure}[h!]
    \centering
    \includegraphics[width=\columnwidth, trim=0 0 0 0, clip]{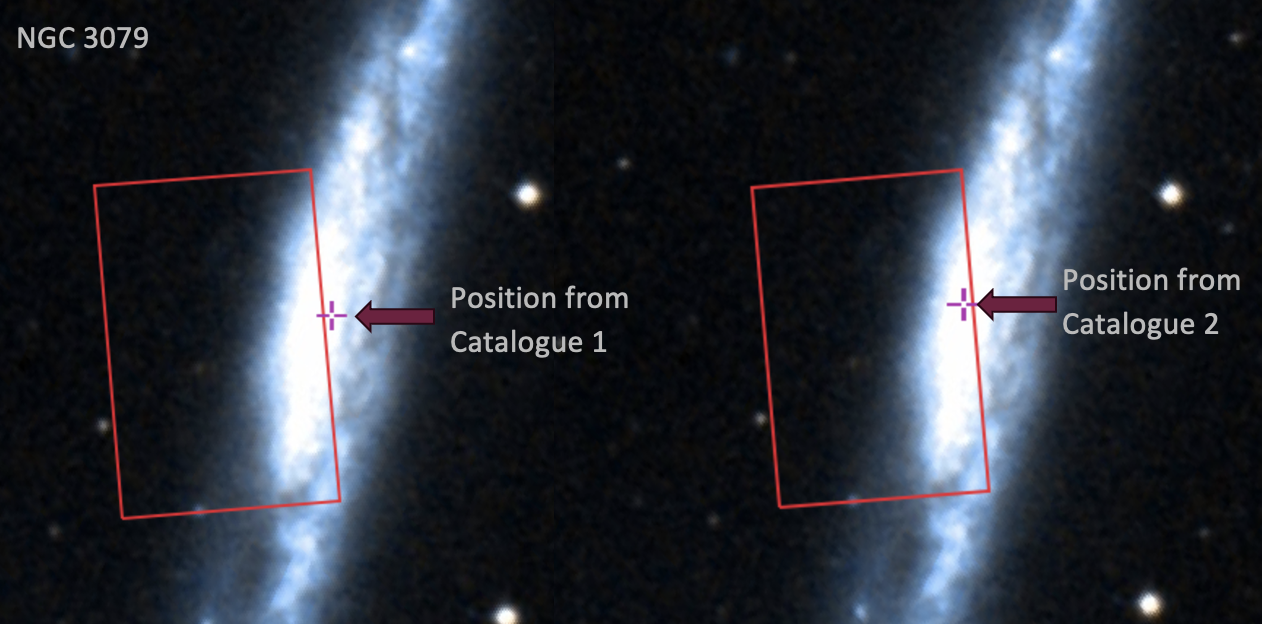}
    \caption{Optical DSS2 image from ESASky of NGC 3079 showing its coordinates (purple cross) according to Swiftbat (left panel) and Milliquas (right panel), and the FoV of observation jw05627017002\_xx102\_00002\_miri (red rectangle). NGC 3079 appears in our catalogue with the Milliquas coordinates.}
    \label{fig:different_pos}
\end{figure} 

The astrometric accuracy of each AGN catalogue (Sect. \ref{Sample}) varies depending on the telescopes used, their resolution, survey epochs, and methodologies used to determine the coordinates of each source. The uncertainties in the coordinates vary not only between catalogues, but sometimes also within a catalogue. Most catalogues report their accuracy for the coordinates for their sources, for example less than $2\arcsec$ for CDFS, Rosat, Stripe,  and \textit{Gaia}, or a few milliarcseconds for MaNGA. Some catalogues, especially the largest ones, make it particularly challenging to determine the uncertainty of the source coordinates \citep[e.g. Milliquas; see][for details]{Flesch2023}. This causes some sources to have slightly different coordinates in different catalogues, even after adjusting all the coordinates to J2000.0. 

Thus, during the coordinate cross-matching process to create our catalogue, some targets may be inside an observation's FoV or outside the FoV depending on the reference catalogue used for the coordinates. These targets are included in our catalogue, but they will appear listed only under the reference catalogue, and the coordinates that fall inside the FoV. An example is shown in Fig. \ref{fig:different_pos} of an MIRI/IMAGE observation (FoV size 74" × 113"; see Table \ref{tab:instrument_distr_all}), of the target NGC 3079. The target coordinates from the Milliquas catalogue (RA = 150.493889, Dec = 55.680556) are inside the observation FoV, whereas the target coordinates from the Swiftbat catalogue (RA = 150.4905, Dec = 55.6799) fall outside. In this case, NGC 3079 will appear in our catalogue with the Milliquas coordinates.

\subsection{Extended sources}
The AGN coordinates in the reference catalogues are usually given for the nucleus of the source. Thus, some observations of extended AGN hosts may be excluded from our catalogue, as the nucleus of the galaxy may not be within the observation FoV (e.g. observations mapping the star formation in the outer regions of an AGN host galaxy, where the galactic nucleus is not included). This was mostly found for nearby objects, since here the resolution enables observations targeting specific parts of the galaxy. We are confident that this issue only affects a small part of the AGN observations, since most of these region-specific observations of an AGN-hosting galaxy are targeted observations, and thus most likely labelled as an AGN by the proposer. Furthermore, most of these observations will include a pointing including the nucleus. Therefore, we expect most, if not all, of these targets to be included in our targeted sample.   

\subsection{Masking in the FoV}
\begin{figure}[h]
    \centering
    \includegraphics[width=\columnwidth, trim=0 0 0 0, clip]{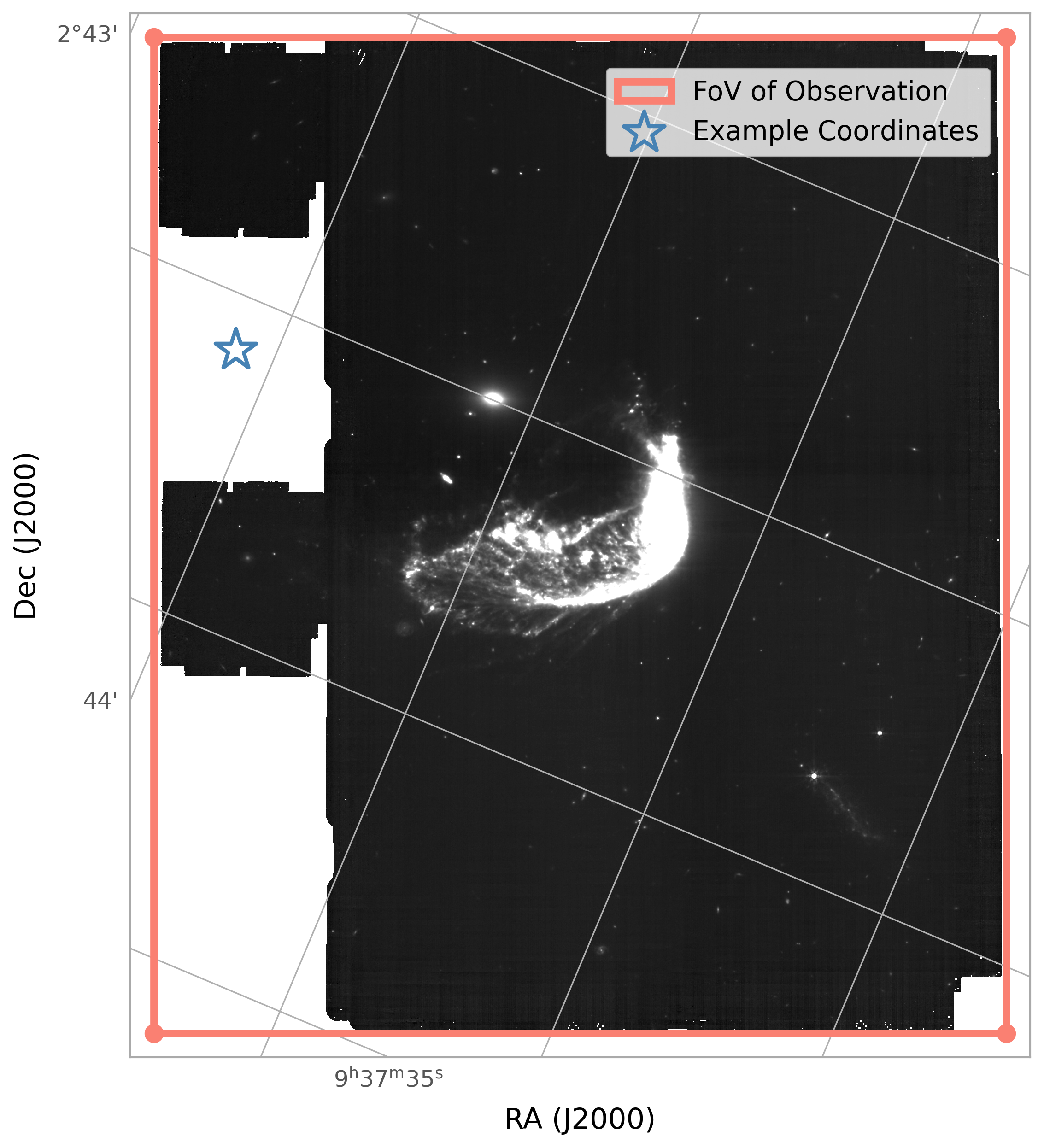}
    \caption{Example of an AGN with coordinates inside the FoV of an observation but in a masked area of the data. This MIRI image of the Penguin Galaxy (NGC 2936) corresponds to observation ID jw06564-o001\_t001\_miri\_f770w. The coloured rectangle shows the FoV as obtained from the position\_bound\_spoly keyword. The blue star shows an example of  coordinates that fall within the FoV but lack data due to masking (in this case caused by the MIRI coronagraphs).}
    \label{fig:masked_area2}
\end{figure}

The polygons describing the observations FoV provided by CAOM cover the whole FoV of the instrument, including regions that may be masked by hardware,  bad pixels, etc. Thus, there could be sources included in our catalogue that fall within
the observations FoV, but
coincide with one of these areas (see Fig. \ref{fig:masked_area2}). 

Currently, we need to locate and remove those observations manually, so there will be cases included in the catalogue. 
Addressing these, and finding a suitable approach is part of future improvements of our catalogue.

\subsection{Planned observations}
Our catalogue includes both observed and planned JWST observations where AGNs fall within the data FoV. 
Over time, especially for proposed observations or pipeline reprocessing, some data details may change. New versions of the catalogue will account for those changes.

 \begin{figure}[ht]
    \centering
    \hspace{2.3cm} 
 \includegraphics[width=1.0\columnwidth]{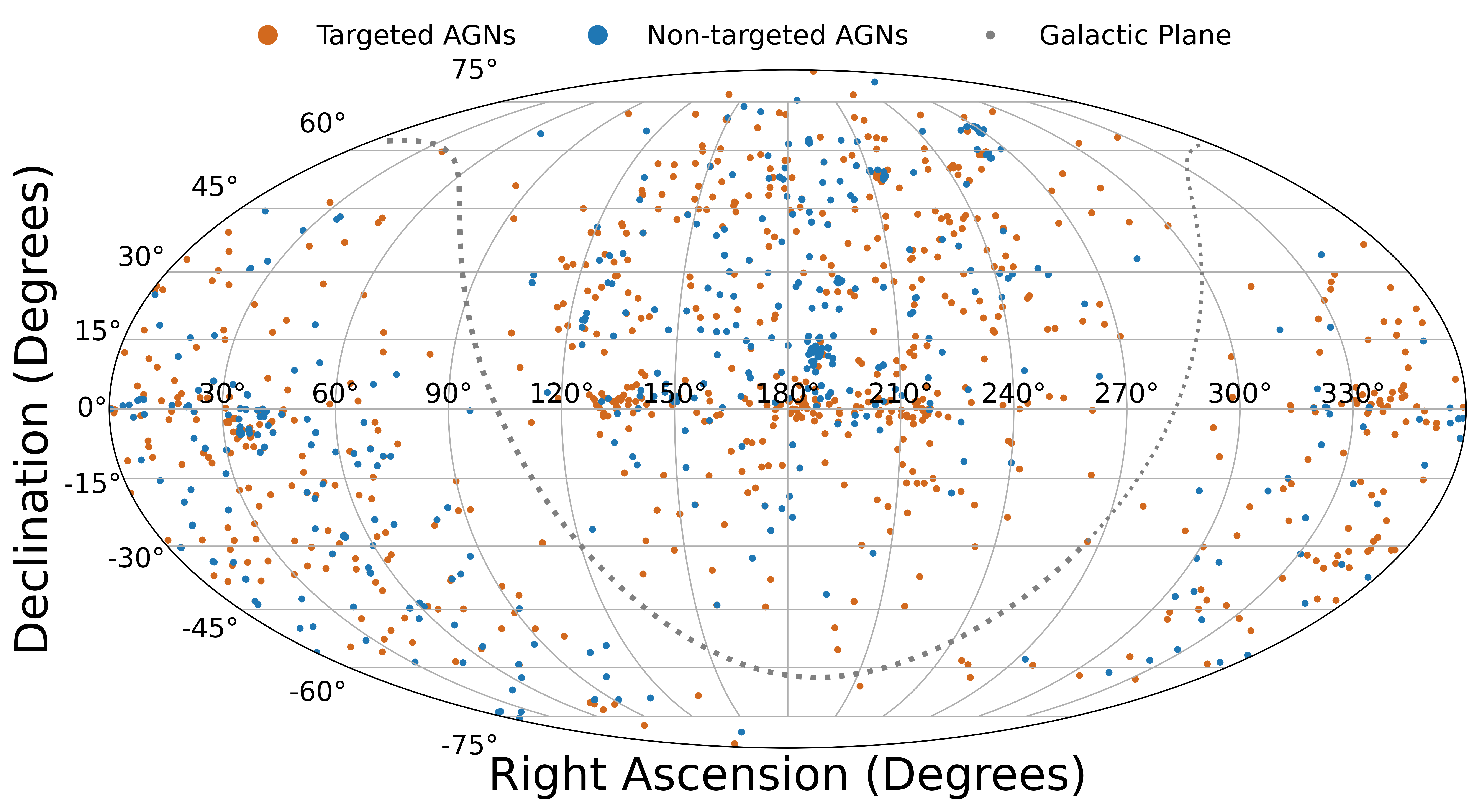}
    \caption{AGNs in the catalogue. This plot includes targeted and non-targeted AGNs in a Mollweide projection. Orange and blue dots mark the coordinates of targeted and non-targeted AGNs (respectively) included in the catalogue. The dotted line follows the Galactic plane in the sky.}
    \label{fig:sky_distribution}
\end{figure}
 
\section{AGN surveys and little red dots}
\label{ambigious AGNs}

\subsection{On-going AGN surveys} 
JWST has enabled a wide range of observational surveys, many of which are expected to uncover new AGNs. These programmes span diverse objectives, through extragalactic surveys and targeted investigations covering different types of sources, from the local to the distant Universe.

Extragalactic surveys such as the JWST Advanced Deep Extragalactic Survey (JADES) \citep[PIDs 1180, 1181, 1210, 1286 1895, 1963, 3215,][]{Eisenstein2023}, the Cosmic Evolution Early Release Science Survey (CEERS) \citep[PID 1345,][]{Finkelstein2017}, and the Systematic Mid-infrared Instrument (MIRI) Legacy Extragalactic Survey (SMILES) \citep[PID 1207;][]{Rieke2024} are designed to explore the high-redshift Universe and are expected to discover many high-redshift AGNs. Targeted programmes such as FRESCO \citep[PID 1895;][]{Oesch2021} and COSMOS-Web \citep[PID 1727;][]{Casey2023} concentrate on specific regions of the sky, with FRESCO observing the GOODS legacy fields and COSMOS-Web extending studies of the COSMOS field. The AGNs discovered by these and other similar JWST programmes have been reported in several studies \citep{Kocevski2023, Scholtz2023, Larson2023, Carniani2024, Matthee2024, Maiolino2024, Ubler2024, Kocevski2024, Lyu2024, Juod2024}.

In addition to large-scale extragalactic efforts, JWST programmes also investigate AGNs in the local and intermediate-redshift Universe. Notable examples include the ERS Q3D programme \citep[PIDs 1335 and 3807;][]{Wylezalek2022, Bertemes2024} and GATOS \citep[PIDs 1670, 2064, 4225, 3535, 4892, 5017, 7195, 7802, and 7429;][]{ GarciaBernete2024, GarciaBernete2024b, ZhangPackham2024}, as well as studies targeting (ultra-)luminous infrared galaxies, for example programmes such as PIDs 1204 and 3368, GOALS \citep{Rich2023}, and individual work by \citet{Huang2023}and \citet{Yui2024}. These objects, among the brightest in the local Universe, often host heavily obscured AGNs.

Most of these surveys are ongoing and collecting or analysing data; therefore, their discoveries of AGNs are still in progress.
Some publications have already listed AGNs or AGN candidates identified in the above-mentioned surveys \citep[e.g. AGNs found in SMILES using the MIRI instrument;][]{Lyu2024}. However, due to the ongoing nature of these programmes, we decided not to include such listings until more complete catalogues are published.
Some AGNs from these surveys are included in our catalogue, if their JWST observations were labelled as AGNs in the proposal or if matches were found in the catalogues used for cross-matching. For example, observations from the JADES survey under PIDs 1180 and 1181 were not explicitly labelled as AGNs but were included in our catalogue through the cross-matching process. In contrast, PID 1670 from the GATOS programme provided explicit AGN labels for their observations, and thus they were categorized as targeted observations in our catalogue. This approach is consistent across the other listed surveys.

In the future, as survey-specific AGN catalogues become available, we plan to incorporate them into ours following the same methodology described above.

\subsection{Little red dots}
With the beginning of JWST observations, previously hidden objects have been uncovered. Among these discoveries, there is a new population of compact, red galaxies identified in the first observations of the JWST CEERS programme \citep{Yang2023, Labbe2023}, known as little red dots (LRDs). These galaxies are characterized by their bimodal spectral energy distributions, exhibiting both a Lyman break and a Balmer break \citep{Labbe2023, Barro2024}.
The classification of LRDs remains an open question, with ongoing debates about whether they should be categorized as AGNs \citep{Barro2024, Matthee2024, Kocevski2024, Akins2024}. Some studies suggest that the spectral energy distributions of LRDs are consistent with significant AGN contributions \citep{Durodola2024}, and the detection of very broad Balmer emission lines \citep{Greene2024} further supports this interpretation. However, other reports highlight properties of LRDs that are difficult to explain with an AGN classification. For instance, the absence of X-ray detections \citep{Yue2024, Tonima2024, Kokubo2024} and the lack of variability \citep{Kokubo2024}, both typically associated with AGNs, challenge this scenario. Additionally, studies show that for some LRDs the broad-line signatures are found to be in agreement with behaviour found in galaxies that do not host an AGN \citep{Baggen2024, Kokubo2024, PerezGonzales2024}.

Because these objects are relatively new, they are not included in the external catalogues we used, and the eJWST proposals do not provide a dedicated LRD keyword. Collecting all currently published LRDs without an existing comprehensive catalogue is beyond the scope of the present work. Therefore, we have not included LRDs in our catalogue. However, if future research provides a comprehensive catalogue of these objects, we will incorporate them into our sample and flag them as AGN candidates when a consensus on the nature of these sources is reached.

\section{Catalogue description}
\label{catalogue description}
Our catalogue contains 3,242 AGNs (968 targeted AGNs and 2,274 non-targeted AGNs), observed in a total of 360,260 calibration level 3 (science-ready) observations plus 47,650 planned observations (including those from Cycle 4 proposals). Among the level 3 observations, 267,219 are public, while 93,041 remain private. Breaking down the observation types for the level 3 observations, 329,425 are classified as science, 1,000 as background, and 1,540 as calibration observations.

The catalogue is structured as a comma-separated value table and includes various columns: Name, Coordinates (RA and Dec), ObjType, Ref, z, zType, zRef, zNED, Observation ID, Instrument, Aperture, Exptype, Filter, Calibration Level, Exptime, Public, Proposal ID, Release Date, Intent, Central Wavelength, Targeted, Data Link, Proposal ID Link, TSOVISIT, eJWST Name, and eJWST Coordinates, as well as Name, ID, z, zType, zRef, and objType for all external catalogues used (see Tables \ref{tab:column_descriptions} and \ref{tab:subset}).
For each AGN, the catalogue lists the Observation ID corresponding to different observations identified through our cross-matching procedure, along with object names. These names may be sourced from the external catalogues, JWST proposals, or generated from the J2000 coordinates. The coordinates are provided in degrees (J2000). Additionally, the catalogue includes direct download links for data files (when publicly available) as well as links to the original proposals.
We include a flag indicating which sources have time-series observations (TSOs) with JWST, as reported in the JWST archive via the TSOVISIT keyword. Sources reported as TSOs are flagged with `t', while those reported as non-TSOs are marked with `n'. Sources without a TSOVISIT entry in the archive are marked as `NaN'. 

\begin{table*}[h!]
\centering
\caption{Comparison of the number of all observations of every AGN per instrument mode in the catalogue.}
\begin{tabular}{lrrr}
\toprule
Instrument & Targeted & Non-Targeted & FoV (arcsec)\\
\midrule
{NIRCAM} & & & \\
\hspace{0.3cm} GRISM & 17,183 & 160,204 & 129 × 129\\ 
\hspace{0.3cm} IMAGE & 1,015 & 34,430 & 2 × 132 × 132\\ 
\hspace{0.3cm} CORON & 0 & 4 & 20 × 20 \\
\hspace{0.3cm} TARGACQ & 0 & 2 & \\
\midrule
{MIRI} & & & \\
\hspace{0.3cm} IFU & 2,112 & 485 & 3.7 × 3.7 to 7.7 × 7.7 \\
\hspace{0.3cm} IMAGE & 1,430 & 8,310 & 74 × 113 \\
\hspace{0.3cm} SLIT & 103 & 11 & 0.51 × 4.7 \\
\midrule
{NIRISS} & & & \\
\hspace{0.3cm} WFSS & 408 & 97,085 & 133 × 133 \\
\hspace{0.3cm} IMAGE & 89 & 520 & 133 × 133 \\
\hspace{0.3cm} AMI & 15 & 27 & 5.2 × 5.2 \\
\midrule
{NIRSpec} & & & \\
\hspace{0.3cm} IFU & 2,047 & 172 & 3.0 × 3.0\\
\hspace{0.3cm} SLIT & 688 & 7 & 0.2 × 3.2, 0.4 × 3.65, 1.6 × 1.6 \\
\hspace{0.3cm} MSA & 28 & 81,559 & 216 × 204 \\
\midrule
{FGS} & & & \\
\hspace{0.3cm} FGS1 & 0 & 2 & 138 × 138 \\
\hspace{0.3cm} FGS2 & 0 & 1 & 138 × 138 \\
\bottomrule
\end{tabular}
\caption*{This table includes both science-ready and planned observations. Notice that individual AGNs can have multiple observations. The table also includes the FoV in arcseconds and the point spread function (PSF) full width at half maximum (FWHM) in arcseconds for each instrument observing mode. For the NIRSpec Microshutter Array (MSA) FoV, the dimensions represent the individual shutter size in the dispersion direction and spatial direction. The PSF FWHM values reflect a range due to their dependence on the different filters and channels used.}
\label{tab:instrument_distr_all}
\end{table*}
We note that this catalogue does not determine or analyse whether an observation is a TSO, it simply reports the information provided by the eJWST archives.
In addition, we incorporated the matches from the external catalogues, preserving the name, coordinates, and redshift columns for each catalogue used. As a result, if a target appears in multiple external catalogues, this will be reflected in the corresponding columns for those catalogues.
A full description of all columns can be found in Table \ref{tab:column_descriptions}.

We list redshift information for a total of 2,814 AGNs, taken from  either the external catalogues (for non-targeted AGNs) or NED (for targeted and non-targeted AGNs). We show the redshift distribution of all targets in the catalogue in Fig. \ref{fig:redshift}. We notice that the non-targeted AGNs predominantly have redshifts below $z = 3$, possibly reflecting the observational limits of previous telescopes. In contrast, the targeted AGNs tend to be higher-redshift sources, highlighting the current interest in early Universe AGNs and showcasing JWST's capabilities to observe them.

Table \ref{tab:instrument_distr_all} compares the distribution of AGNs in the catalogue per instrument and observing mode.
Most observations correspond to the NIRCam/GRISM mode, for both targeted and non-targeted AGNs. Additionally, for targeted AGNs, there is a substantial number of IFU observations (MIRI and NIRSpec). This preference likely stems from the unique advantages of IFUs in providing spatially resolved spectral information.
In contrast, a significant number of non-targeted AGNs have been identified using the NIRISS/WFSS and NIRSpec/Microshutter Array instruments. This is probably due to their large FoVs, which allow them to include more sources in their observations. 

In Fig. \ref{fig:sky_distribution} we present the sky distribution of all targeted and non-targeted AGNs from our study using a Mollweide projection. The broad distribution of positions demonstrates that our external catalogue search did not preferentially focus on specific regions of the sky. It also highlights how the observing programmes tend to avoid the galactic plane, as expected.

\section{Conclusion}
\label{conclusion}
In this work we present a catalogue of all AGNs observed with JWST, whether as direct targets or serendipitous observations in the FoV of any dataset, complete by March 20 2025, right after the publication of the Cycle 4 results. The catalogue lists observations of 3,242 individual AGNs.
For the targeted AGNs, we applied two distinct selection approaches. First, we used ADQL queries to filter AGN observations by specific keywords designated by the proposers. Second, we manually reviewed relevant JWST programme proposals to identify additional targeted AGNs. This combined approach resulted in the identification of 968 individual targeted AGNs. For the non-targeted AGNs, we identified objects located within the JWST FoV by cross-matching multiple external AGN catalogues with JWST observations, ensuring that duplicate entries were systematically removed when necessary. In total, 2,274 non-targeted AGNs were identified through this process.

Observations of nearby or extended hosts of AGNs may be underrepresented in our catalogue if the FoV of any observations does not include the nuclear region, as the catalogues used for the cross-correlation typically give the coordinates of the nucleus.
We expect such omissions to be minimal, as close-up observations of AGN hosts are likely captured by our targeted AGN selection. Additionally, it is possible that an AGN falls within the FoV of an observation but lies in a masked region of the instrument, resulting in no usable data for that target. Currently, we are working on automatizing a comprehensive method for accounting for these cases, 
as well as the (currently manual) proposal review process to identify targeted AGNs. 

For future releases, we aim to update the catalogue at least once a year to include new observed and planned data, and incorporate additional AGN catalogues and surveys, as well as LRDs and potentially new sources, as data become available. 

\section*{Data availability}
The last update of our catalogue, at the time of writing, was completed on March 20, 2025, to include the Cycle 4 approved proposals. As the archives are constantly updated with new data, we aim to update our catalogue on an --- at least --- yearly basis, in time for every JWST call for proposals.

Code for this study and the latest version of the catalogue is available at GitHub\footnote{\url{https://github.com/VirginiaLenk1/AGNs-in-eJWST.git}}.
The catalogue presented in this work (Table \ref{tab:subset}) is available in electronic form at the CDS via anonymous FTP (ftp://cdsarc.u-strasbg.fr) or via http://cdsweb.u-strasbg.fr/cgi-bin/qcat?J/A+A/.

\begin{acknowledgements}
We thank the referee for their suggestions that helped improve the catalogue and the paper.
VL thanks the SCI-S Trainee Programme of the European Space Agency (ESA) for providing me with the incredible opportunity to work on this project. The support and resources provided by the program were instrumental in the successful completion of this work. Additional thanks to M. Arévalo and J. Espinosa for their valuable assistance in navigating the eJWST archive and their guidance to CAOM.
CC acknowledges support from the ESA Fellowship in Space Science program.
AL acknowledges support from ESA through the ESA Space Science Faculty support programs.
AAH acknowledges support from grant PID2021-124665NB-I00 funded by the Spanish
Ministry of Science and Innovation and the State Agency of Research
MCIN/AEI/10.13039/501100011033  and ERDF A way of making Europe.
The authors thank AGN group at ESAC, and its visitors, for useful scientific discussions and feedback on how to improve the catalogue and source selection.
This research has made use of NASA's Astrophysics Data System Bibliographic Services, and of the NASA/IPAC Extragalactic Database (NED), which is operated by the Jet Propulsion Laboratory, California Institute of Technology, under contract with the National Aeronautics and Space Administration.
This research has made use of the Spanish Virtual Observatory (https://svo.cab.inta-csic.es) project funded by MCIN/AEI/10.13039/501100011033/ through grant PID2020-112949GB-I00.
This work is based on observations made with the NASA/ESA/CSA \textit{James Webb} Space Telescope. The data were obtained from the \href{https://jwst.esac.esa.int/archive/}{European JWST archive (e{JWST})}, which is operated by the European Space Agency; and the Mikulski Archive for Space Telescopes at the Space Telescope Science Institute, which is operated by the Association of Universities for Research in Astronomy, Inc., under NASA contract NAS 5-03127 for JWST. 
\end{acknowledgements}
\bibliographystyle{aa} 
\bibliography{Bibliography}

\appendix
\onecolumn
\section{Object type mapping across catalogues} \label{app:objtype}
For consistency across catalogues, the object type flags were mapped to a unified scheme. Tables~\ref{tab:milliquas_map}--\ref{tab:stripe_map} summarize all mappings used in the construction of the catalogue, where Original Code shows the classification as listed in the respective catalogues, Unified Type gives the corresponding mapping adopted in our catalogue, and Notes provide additional information from the original published catalogues regarding the abbreviations. Catalogues that did not provide objTypes (e.g. eFEDS, Swiftbat, and CDFS) are not listed here (see Sect. \ref{objtype} for details).

\begin{table}[ht]
\centering
\caption{Milliquas object type mapping.}
\label{tab:milliquas_map}
\begin{tabular}{lp{3cm}p{12cm}}
\hline
Original code & Unified type & Notes \\
\hline
Q & QSO type1-BL & QSO, type-I broad-line core-dominated \\
A & AGN type1    & AGN, type-I Seyferts/host-dominated \\
B & BL Lac       & BL Lac type object \\
K & QSO-NL       & NLQSO, type-II narrow-line core-dominated \\
N & AGN-NL       & NLAGN, type-II Seyferts/host-dominated. Includes an unquantified residue of legacy NELGs/ELGs/LINERs, plus some unclear AGN. \\
S & Star         & Star classified but showing quasar-like photometry and radio/X-ray association, thus included as a quasar candidate \\
R & R            & Radio association flag \\
X & X            & X-ray association flag \\
2 & 2            & Double radio lobes displayed \\
\hline
\end{tabular}
\end{table}

\begin{table}[ht]
\centering
\caption{Veroncat object type mapping.}
\label{tab:veroncat_map}
\begin{tabular}{lll}
\hline
Original code & Unified type & Notes \\
\hline
Q2   & QSO type2              & Type-2 QSO \\
S1   & Seyfert 1              & Seyfert 1 spectrum \\
S1h  & Seyfert 1 --h          & Broad polarized Balmer lines detected \\
S1i  & Seyfert 1 --i          & Broad Paschen lines observed in the infrared \\
S1n  & Seyfert 1-NL           & Narrow-line Seyfert 1 \\
S1.0 & Seyfert 1.0            & \\
S1.2 & Seyfert 1.2            & \\
S1.5 & Seyfert 1.5            & \\
S1.8 & Seyfert 1.8            & \\
S1.9 & Seyfert 1.9            & \\
S2   & Seyfert 2              & Seyfert 2 spectrum \\
S2?  & Seyfert 2 ?            & Probable Seyfert 2 \\
S3   & Seyfert 3/LINER      & Seyfert 3 or liner\\
S3b  & Seyfert 3/LINER --b    & Seyfert 3 or liner with broad Balmer lines\\
S3h  & Seyfert 3/LINER --h    & Seyfert 3 or liner with broad polarized Balmer lines detected \\
S    & Seyfert (unclassified) & unclassified Seyfert \\
S?   & Seyfert ?              & Probable Seyfert \\
H2   & Nuclear HII region     &  nuclear HII region \\
HP   & High polarization      & high optical polarization (>3\%) \\
BL   & BL Lac                 & Confirmed BL Lac \\
BL?  & BL Lac ?               & Probable BL Lac \\
?    & BL Lac ??              & Questionable BL Lac \\
\hline
\end{tabular}
\end{table}

\begin{table}[ht]
\centering
\caption{MaNGA object type mapping.}
\label{tab:manga_map}
\begin{tabular}{l l p{7cm}}
\hline
Original code & Unified type & Notes \\
\hline
BROAD\_AGN                    & BROAD\_AGN                    & AGN identified by broad lines \\
BAT\_AGN                       & BAT\_AGN                       & AGN identified in \textit{Swift}/BAT hard X-ray survey \\
WISE\_AGN                      & WISE\_AGN                      & AGN identified via WISE mid-IR colours \\
RADIO\_AGN (HERG)              & HERG (quasar mode)            & quasar-mode high-excitation radio galaxy \\
RADIO\_AGN (LERG)              & LERG (radio mode)             & radio-mode low-excitation radio galaxy  \\
\\
\hline
\end{tabular}
\end{table}

\begin{table}[ht]
\centering
\caption{Rosat object type mapping.}
\label{tab:rosat_map} 
\begin{tabular}{l l p{7cm}}
\hline
Original code & Unified type & Notes \\
\hline
QSO & QSO & object with broad lines and $M_B<-22.5$ \\
A & AGN-NL & narrow line objects with [OIII]/H$\beta>2.0$ and/or [NII]/H$\alpha>0.6$ \\
BA & AGN-BL & objects with broad lines and $M_B>-22.5$ \\
H & Starburst & starbursts with [OIII]/H$\beta<2.0$ and/or [NII]/H$\alpha<0.6$ \\
G & Galaxy & galaxies (objects with no emission lines and Ca II break contrasts $>30\%$) \\
B & BL Lac & BL Lacs \\
S & Star & \\
Cl & Cluster candidate & possible clusters (based on the proximity of a known cluster) \\
rad & Radio source & Catalogued radio source \\
Sy & Seyfert & \\
Irs & Infrared source & Identified via IR emission \\
Vis & Optical source & Optical/Visible classification \\
UvE & UV-excess source & UV-excess detection \\
\hline
\end{tabular}
\end{table}

\begin{table}[ht]
\centering
\caption{BASS object type mapping.}
\label{tab:bass_map}
\begin{tabular}{lp{5cm}p{10cm}}
\hline
Original code & Unified type & Notes \\
\hline
Sy1   & Seyfert 1                   & AGN with broad H$\beta$ \\
Sy1.9 & Seyfert 1.9                 & AGN with narrow H$\beta$ and broad H$\alpha$ \\
Sy2   & Seyfert 2                   & AGN with narrow H$\beta$ and H$\alpha$ \\
BZQ   & Blazar -- quasar type       & beamed AGN with the presence of broad lines \\
BZG   & Blazar -- galaxy dominated  & beamed AGN with only Host-galaxy features, no broad lines \\
BZB   & Blazar -- featureless cont. & beamed AGN, traditional continuum dominated blazars with no emission lines or host galaxy features \\
\hline
\end{tabular}
\end{table}

\begin{table}[ht]
\centering
\caption{Stripe object type mapping.}
\label{tab:stripe_map} 
\begin{tabular}{lll}
\hline
Original code & Unified type  \\
\hline
1 & Stars       &  \\
2 & Ellipticals & \\
3 & Spirals     & \\
4 & Type 2      & \\
5 & Starbursts  & \\
6 & Type 1      & \\
7 & QSOs        & \\
\hline
\end{tabular}
\end{table}

\onecolumn
\section{Summary of catalogues}
\begin{table*}[h!]
\centering
\caption{Summary of the catalogues used.}
\begin{tabular}{p{2.8cm} p{1.7cm} p{2.5cm} p{3.6cm} p{3.9cm}}
    \toprule
    {Catalogue Name} & {\# AGNs} & {Redshift Range} & {Area (deg$^2$)} & {Info} \\
    \midrule
    \small Milliquas & \small 1,021,800 & \small 0.001 - 7.642 & \small -- & \small AGNs from different surveys.  \\
    \small MaNGA & \small 398 & \small 0.001 - 0.149 & \small 2,700 (MaNGA Field) & \small Multi-wavelength AGNs.\\
    \small eFEDs & \small 22,079 & \small 0.004 - 8.000 & \small 142 & \small X-ray AGNs. \\
    \small Stripe & \small 2,459 & \small 0.000 - 7.011 & \small 31.3 & \small X-ray AGNs with multi-wavelength counterparts. \\
    \small \textit{Chandra} & \small 711 & \small 0.034 - 4.762 & \small 484.2 & \small X-ray AGNs.\\
    \small BASS & \small 858 & \small 0.001 - 3.656 & \small All-sky (\textit{Swift}/BAT) & \small Hard-X-ray AGNs from VLT.\\
    \small Veroncat & \small 34,231 & \small 0.001 - 5.870 & \small --  & \small AGNs from different surveys. \\
    \small Rosat & \small 294 & \small 0.002 - 4.715 & \small All-sky (Rosat) & \small  X-ray/radio AGNs.\\
    \small \textit{Gaia}-CRF3 &  \small 1,614,173 & \small  -- & \small All-sky (\textit{Gaia}) & \small AGNs in \textit{Gaia} DR3 identified by 17 different catalogues. \\
    \small Swiftbat & \small 277 & \small < 0.06 & \small All-sky (\textit{Swift}/BAT) & \small  X-ray AGNs.\\
    \bottomrule
\end{tabular}
\label{tab:catalog_summary}
\end{table*}

\section{Columns}
\label{columns}
{\renewcommand{\arraystretch}{1.6}
\begin{longtable}{p{4.2cm} p{13cm}}
\caption{Description of the columns.} \label{tab:column_descriptions} \\
\toprule
{Column Nr \& Name} & {Description} \\
\midrule
\endfirsthead
\caption[]{Continued.}\\
\toprule
 {Column Nr \& Name} & {Description} \\
\endhead
(1) Name & Object Names from literature (reference source given in Reference column). Priority was given to information from Milliquas due to completeness. 
Names for CDFS objects were derived from their position in the HHMMSS.SS+DDMMSS.S. format. Names provided by eJWST should be taken with caution, because these are usually introduced by the proposer and typically do not follow a standard convention.  \\ \hline
(2-3) RA/Dec (J2000, deg) & Target coordinates (J2000, in degrees) from literature. Rounded to 7 decimals for display purposes and to avoid truncation. To see real accuracy, go to corresponding catalogue coordinate column (RA (catalogue)/Dec (catalogue)). Coordinates taken from \textit{Gaia} were shifted to J2000. See Sect. \ref{Sample} for more information. \\ \hline
(4) objType & Classification of the source (e.g. Seyfert, QSO, BL Lac, LINER, etc.), retrieved from proposer descriptions for targeted AGNs and from catalogues for non-targeted AGNs. Catalogue-specific codes were harmonized into a unified scheme; `NL' and `BL' indicate narrow-line or broad-line objects, `?' marks uncertain types, and `??' highly questionable classifications. Proposer Descriptions are kept as provided by the eJWST archive. \\ \hline
(5) Ref & Original publication used to determine first four columns. When an object was listed in multiple catalogues, the following priority was used (from left to right): Milliquas - MaNGA - eFEDS - Stripe - CDFS - BASS - Veroncat - Rosat  - eJWST - \textit{Gaia} - Swiftbat. This order was determined by how complete, updated and accurate the information from the individual catalogues were. All catalogues that do not provide reliable (or any) redshifts for the sources were placed close to the bottom. \\ \hline
(6) z & Redshift from literature. Retrieved either from NED or directly from the catalogues used. NED values were given priority. See Sect. \ref{redshift} for more information. Missing redshifts were marked as NaN. \\ \hline
(7) zType & Technique by which the redshift was determined. Information were taken from NED (using its standard convention) or, if unavailable, from the catalogues. NED values were given priority. Catalogue-specific terms were harmonized to a common scheme (spectroscopic = `S', photometric = `P') to be consistent with the NED convention; unknown cases are marked with `U'.  \\ \hline
(8) zRef & Redshift reference in the form of a bibcode for each object. Values were taken from NED when available, otherwise from the respective catalogues. \\ \hline
(9) zNED & Flag indicating whether redshift information was taken from NED (`y') or not (`n'). \\ \hline
(10) Observation ID & Unique identifier for an observation.\\ \hline
(11) Instrument & Name of the instrument used for this observation. \\ \hline
(12) Aperture & Unique targetable fiducial point and its associated region per Instrument\\ \hline
(13) Exptype & Exposure type - type of data in the exposure. \\ \hline
(14) Filter & Bandpass filters in different wavelengths. \\ \hline
(15) Calibrationlevel & Calibration level with Planned observations = -1 and Science-ready observations = 3 \\ \hline
(16) Exptime (s) & Total exposure time. \\ \hline
(17) Public & States if the observation is public or not (True or False). Public = for all available, otherwise restricted to proposer for a certain period of time. \\ \hline
(18) Proposal ID & Collection-specific unique proposal identifier. \\ \hline
(19) Release Date (UTC) & Date the metadata for an observation is public (UTC). \\ \hline
(20) Intent & Intended purpose of observation (Science, Calibration, Background). \\
    & Labelled as follows: \\
    & Science = Scientific observations done with the purpose of targeting a specific object or area in the sky.\\
    & Calibration = Observations used to calibrate the science instruments and observing modes.\\
    & Background = Background observation listed by MAST portal or found through eJWST keywords. \\ \hline
(21) Central Wavelength ($\mu m$) & Central wavelength of the observation. \\ \hline
(22) Targeted & Checks if object was targeted by JWST or not (True or False; see Sect. \ref{agn selection}). \\ \hline 
(23) Data Link & Link for direct data product download when observations are public. \\ \hline
(24) Proposal ID Link & Link to programme information for specific proposal ID.\\ \hline
(25) TSOVISIT & Flag indicating whether an observation is a TSO or not (`t' or `f'). Nan values were assigned for observations not including this keyword on the eJWST archive.
\\ \hline
(26) eJWST Name & Name of object as given by the proposer, listed as target\_name in eJWST. \\ 
(27) RA (eJWST) & Right ascension in deg and J2000, listed as targetposition\_coordinate\_cval1 in eJWST.\\ 
(28) Dec (eJWST) & Declination in deg and J2000, listed as targetposition\_coordinate\_cval2 in eJWST. Note: These coordinates are found in jwst.observation, in jwst.archive the same keywords store the pointing coordinates of the instruments per observation. \\ 
(29) objType (eJWST)  & Classification of object as listed by proposer in the eJWST archive.\\ 
(30) z (NED) & Redshift taken from the astronomical database NED by a coordinate search with 1 arcsec radius.\\
(31) zType (NED) & Technique used to determine the redshift. Information taken from the astronomical database NED by a coordinate search with 1 arcsec radius. \\
(32) zRef (NED) & Redshift Reference taken from the astronomical database NED by a coordinate search with 1 arcsec radius. \\ \hline
Name ($^*$) & Name of object taken from ($^*$) catalogue. \\
ID ($^*$) & Source ID in  ($^*$) catalogue, when available. \\
RA ($^*$) & Right Ascension in decimal degrees (J2000) taken from  ($^*$) catalogue.\\ 
Dec ($^*$) & Declination in decimal degrees (J2000) taken from  ($^*$) catalogue.\\
z ($^*$) &  Redshift taken from  ($^*$) catalogue, when available. \\
zType ($^*$) & Technique used to determine the redshift. Information taken from ($^*$) catalogue, when available. Spectroscopic redshift was labelled `spec', photometric `phot' and unknown cases `?'.  \\ 
zRef ($^*$) & Redshift Reference of object taken from ($^*$) catalogue, when available.  \\ 
objType ($^*$) & Classification of object taken from ($^*$) catalogue, when available.  \\ 
\bottomrule
\end{longtable}}
\footnotesize{
\noindent
{($^*$)} These fields are given for different catalogues. Depending on the AGN, they will correspond to one or several of the following: 
\begin{itemize}[label=$\bullet$]
    \item (33–41) {Milliquas} by \citet{Flesch2023}. Redshift type: spectroscopic with a precision better than 0.1; photometric when rounded to 0.1. Extra columns: XName (36) / RName (37) – Additional names from X-ray/Radio detections listed in Milliquas.
    \item (42–48) {MaNGA} by \citet{Comerford2024}. Redshift type: spectroscopic. 
    \item (49–55) {eFEDS} by \citet{Lui2022}. Redshift type: from `CTP\_REDSHIFT\_GRADE', with 5 = spectroscopic, 2–4 = photometric.
    \item (56–63) {Stripe} by \citet{Ananna2019}. Name of object derived from their position in `JHHMMSS.SS+DDMMSS.SS' format. Redshift type: spectroscopic. Extra columns: ID (Stripe) (57) - ID number for each source as assigned in \citet{Lamassa2016}.
    \item (64–71) {CDFS} by \citet{Luo2017}. Name of object derived from their position in `JHHMMSS.SS+DDMMSS.SS' format. Redshift type: spectroscopic. Extra columns: ID (CDFS) (65) – X-ray Source ID.
    \item (72–79) {BASS} by \citet{Koss2022}. Redshift type: spectroscopic when emission lines used (e.g. [O III], H$\alpha$, Mg II, and CIV); `?' otherwise. Extra columns: ID (BASS) (73) – catalogue ID in the BASS DR2 survey.
    \item (80–86) {Veroncat} by \citet{VCV2010}. Redshift type: spectroscopic if `l\_z' column flagged by `*'; `?' otherwise.
    \item (87–93) {Rosat} by \citet{Brinkmann2000}. Redshift type: unknown, flagged as `?'. 
    \item (94–96) {\textit{Gaia}} by \citet{Gaia2023}. Coordinates transformed from J2016 coordinates. No redshifts.
    \item (97–99) {Swiftbat} by \citet{Lutz2018L}. No redshifts.
\end{itemize}
}
\twocolumn

\begin{sidewaystable*}[t]
  \centering
  \caption{Sample lines of our JWST AGN catalogue.}
  \resizebox{\textwidth}{!}{%
\begin{tabular}{lrrlllllllll}
\toprule
Name & RA (J2000, deg) & Dec (J2000, deg) & objType & Ref & z & zType & zRef & zNED & Observation ID & Instrument & Aperture \\
\midrule
NGC 4388 & 186.444683 & 12.661875 & AGN-NL-R-X & Millquas & 0.008419 & SUN & 1993ApJS...88..383L & y & jw09218003001\_xx101\_00001\_miri & MIRI/IMAGE & MIRIM\_SUB256 \\
NGC 4388 & 186.444683 & 12.661875 & AGN-NL-R-X & Millquas & 0.008419 & SUN & 1993ApJS...88..383L & y & jw09218003001\_xx102\_00002\_miri & MIRI/IMAGE & MIRIM\_SUB256 \\
NGC 4388 & 186.444683 & 12.661875 & AGN-NL-R-X & Millquas & 0.008419 & SUN & 1993ApJS...88..383L & y & jw09218003001\_xx104\_00004\_miri & MIRI/IMAGE & MIRIM\_SUB256 \\
NGC 4388 & 186.444683 & 12.661875 & AGN-NL-R-X & Millquas & 0.008419 & SUN & 1993ApJS...88..383L & y & jw09218003001\_xx105\_00001\_miri & MIRI/IMAGE & MIRIM\_SUB256 \\
NGC 4388 & 186.444683 & 12.661875 & AGN-NL-R-X & Millquas & 0.008419 & SUN & 1993ApJS...88..383L & y & jw09218003001\_xx106\_00002\_miri & MIRI/IMAGE & MIRIM\_SUB256 \\
NGC 4388 & 186.444683 & 12.661875 & AGN-NL-R-X & Millquas & 0.008419 & SUN & 1993ApJS...88..383L & y & jw09218003001\_xx108\_00004\_miri & MIRI/IMAGE & MIRIM\_SUB256 \\
NGC 4388 & 186.444683 & 12.661875 & AGN-NL-R-X & Millquas & 0.008419 & SUN & 1993ApJS...88..383L & y & jw09218003001\_xx109\_00001\_miri & MIRI/IMAGE & MIRIM\_SUB256 \\
NGC 4388 & 186.444683 & 12.661875 & AGN-NL-R-X & Millquas & 0.008419 & SUN & 1993ApJS...88..383L & y & jw09218003001\_xx10a\_00002\_miri & MIRI/IMAGE & MIRIM\_SUB256 \\
NGC 4388 & 186.444683 & 12.661875 & AGN-NL-R-X & Millquas & 0.008419 & SUN & 1993ApJS...88..383L & y & jw09218003001\_xx10c\_00004\_miri & MIRI/IMAGE & MIRIM\_SUB256 \\
NGC 4395 & 186.453600 & 33.546860 & AGN-NL-R-X & Millquas & 0.001064 & SUN & 1998AJ....115...62H & y & jw02016-c1007\_t002\_miri\_ch1-shortmediumlong & MIRI/IFU & MIRIFU\_CHANNEL1A \\
NGC 4395 & 186.453600 & 33.546860 & AGN-NL-R-X & Millquas & 0.001064 & SUN & 1998AJ....115...62H & y & jw02016-c1007\_t002\_miri\_ch2-shortmediumlong & MIRI/IFU & MIRIFU\_CHANNEL1A \\
NGC 4395 & 186.453600 & 33.546860 & AGN-NL-R-X & Millquas & 0.001064 & SUN & 1998AJ....115...62H & y & jw02016-c1007\_t002\_miri\_ch3-shortmediumlong & MIRI/IFU & MIRIFU\_CHANNEL1A \\
NGC 4395 & 186.453600 & 33.546860 & AGN-NL-R-X & Millquas & 0.001064 & SUN & 1998AJ....115...62H & y & jw02016-c1007\_t002\_miri\_ch4-shortmediumlong & MIRI/IFU & MIRIFU\_CHANNEL1A \\
NGC 4395 & 186.453600 & 33.546860 & AGN-NL-R-X & Millquas & 0.001064 & SUN & 1998AJ....115...62H & y & jw02016-o002\_t002\_miri\_ch1-shortmediumlong & MIRI/IFU & MIRIFU\_CHANNEL1A \\
NGC 4395 & 186.453600 & 33.546860 & AGN-NL-R-X & Millquas & 0.001064 & SUN & 1998AJ....115...62H & y & jw02016-o002\_t002\_miri\_ch2-shortmediumlong & MIRI/IFU & MIRIFU\_CHANNEL1A \\
NGC 4395 & 186.453600 & 33.546860 & AGN-NL-R-X & Millquas & 0.001064 & SUN & 1998AJ....115...62H & y & jw02016-o002\_t002\_miri\_ch3-shortmediumlong & MIRI/IFU & MIRIFU\_CHANNEL1A \\
NGC 4395 & 186.453600 & 33.546860 & AGN-NL-R-X & Millquas & 0.001064 & SUN & 1998AJ....115...62H & y & jw02016-o002\_t002\_miri\_ch4-shortmediumlong & MIRI/IFU & MIRIFU\_CHANNEL1A \\
NGC 4395 & 186.453600 & 33.546860 & AGN-NL-R-X & Millquas & 0.001064 & SUN & 1998AJ....115...62H & y & jw02016-o028\_t001\_nirspec\_g235h-f170lp & NIRSPEC/IFU & NRS\_FULL\_IFU \\
NGC 4395 & 186.453600 & 33.546860 & AGN-NL-R-X & Millquas & 0.001064 & SUN & 1998AJ....115...62H & y & jw02016-o028\_t001\_nirspec\_g395h-f290lp & NIRSPEC/IFU & NRS\_FULL\_IFU \\
NGC 4418 & 186.727583 & -0.877612 & AGN-NL-R-X & Millquas & 0.007085 & SLS & 2016SDSSD.C...0000: & y & jw01991-c1003\_t003\_miri\_ch1-shortmediumlong & MIRI/IFU & MIRIFU\_CHANNEL1A \\
NGC 4418 & 186.727583 & -0.877612 & AGN-NL-R-X & Millquas & 0.007085 & SLS & 2016SDSSD.C...0000: & y & jw01991-c1003\_t003\_miri\_ch2-shortmediumlong & MIRI/IFU & MIRIFU\_CHANNEL1A \\
NGC 4418 & 186.727583 & -0.877612 & AGN-NL-R-X & Millquas & 0.007085 & SLS & 2016SDSSD.C...0000: & y & jw01991-c1003\_t003\_miri\_ch3-shortmediumlong & MIRI/IFU & MIRIFU\_CHANNEL1A \\
NGC 4418 & 186.727583 & -0.877612 & AGN-NL-R-X & Millquas & 0.007085 & SLS & 2016SDSSD.C...0000: & y & jw01991-c1003\_t003\_miri\_ch4-shortmediumlong & MIRI/IFU & MIRIFU\_CHANNEL1A \\
NGC 4418 & 186.727583 & -0.877612 & AGN-NL-R-X & Millquas & 0.007085 & SLS & 2016SDSSD.C...0000: & y & jw01991-o006\_t003\_miri\_ch1-shortmediumlong & MIRI/IFU & MIRIFU\_CHANNEL1A \\
NGC 4418 & 186.727583 & -0.877612 & AGN-NL-R-X & Millquas & 0.007085 & SLS & 2016SDSSD.C...0000: & y & jw01991-o006\_t003\_miri\_ch2-shortmediumlong & MIRI/IFU & MIRIFU\_CHANNEL1A \\
NGC 4418 & 186.727583 & -0.877612 & AGN-NL-R-X & Millquas & 0.007085 & SLS & 2016SDSSD.C...0000: & y & jw01991-o006\_t003\_miri\_ch3-shortmediumlong & MIRI/IFU & MIRIFU\_CHANNEL1A \\
NGC 4418 & 186.727583 & -0.877612 & AGN-NL-R-X & Millquas & 0.007085 & SLS & 2016SDSSD.C...0000: & y & jw01991-o006\_t003\_miri\_ch4-shortmediumlong & MIRI/IFU & MIRIFU\_CHANNEL1A \\
NGC 4418 & 186.727583 & -0.877612 & AGN-NL-R-X & Millquas & 0.007085 & SLS & 2016SDSSD.C...0000: & y & jw01991-o007\_t004\_miri\_f1000w-sub256 & MIRI/IMAGE & MIRIFU\_CHANNEL1A \\
NGC 4418 & 186.727583 & -0.877612 & AGN-NL-R-X & Millquas & 0.007085 & SLS & 2016SDSSD.C...0000: & y & jw01991-o007\_t004\_miri\_f1130w-sub256 & MIRI/IMAGE & MIRIFU\_CHANNEL1A \\
NGC 4418 & 186.727583 & -0.877612 & AGN-NL-R-X & Millquas & 0.007085 & SLS & 2016SDSSD.C...0000: & y & jw01991-o007\_t004\_miri\_f770w-sub256 & MIRI/IMAGE & MIRIFU\_CHANNEL1A \\
NGC 4418 & 186.727583 & -0.877612 & AGN-NL-R-X & Millquas & 0.007085 & SLS & 2016SDSSD.C...0000: & y & jw01991-o008\_t006\_nirspec\_g395m-f290lp & NIRSPEC/IFU & NRS\_FULL\_IFU \\
NGC 4438 & 186.940000 & 13.008889 & Seyfert 3/Liner -b & Veroncat & 0.000237 & SUN & 1995ApJ...438..135K & y & jw07763006001\_xx101\_00001\_nircam & NIRCAM/IMAGE & NRCALL\_FULL \\
NGC 4438 & 186.940000 & 13.008889 & Seyfert 3/Liner -b & Veroncat & 0.000237 & SUN & 1995ApJ...438..135K & y & jw07763006001\_xx102\_00002\_nircam & NIRCAM/IMAGE & NRCALL\_FULL \\
NGC 4438 & 186.940000 & 13.008889 & Seyfert 3/Liner -b & Veroncat & 0.000237 & SUN & 1995ApJ...438..135K & y & jw07763006001\_xx103\_00003\_nircam & NIRCAM/IMAGE & NRCALL\_FULL \\
NGC 4438 & 186.940000 & 13.008889 & Seyfert 3/Liner -b & Veroncat & 0.000237 & SUN & 1995ApJ...438..135K & y & jw07763006001\_xx104\_00004\_nircam & NIRCAM/IMAGE & NRCALL\_FULL \\
\bottomrule
\end{tabular}
  }
  \tablefoot{Column (1): Name adopted from literature. Column (2-3): Right Ascension and
    Declination in decimal degrees, based on J2000 epoch. Column (4): Object Type
    adopted from literature and the eJWST archive. Column (5): Source of information
    for columns 1-4. Column (6): Redshift, taken either from NED or from the same
    source as in column 5. Column (7): Redshift type/method. Column (8): Redshift
    reference. Column (9): NED flag, indicating whether the redshift was taken from
    NED (y) or from the same source as listed in column 5 (n). Column (10): Unique
    identifier for an observation. Columns (11-12): Observation specific information.
    A detailed description of this table’s contents is given in Table
    \ref{tab:column_descriptions}.}
  \label{tab:subset}
\end{sidewaystable*}

\end{document}